\documentclass[english,10pt,letterpaper]{article}
\usepackage[T1]{fontenc}
\usepackage[latin9]{inputenc}
\usepackage{amsbsy}
\usepackage{amstext}
\usepackage{amssymb}
\usepackage{amsmath}
\usepackage{graphicx}
\usepackage{esint}
\usepackage{bm}

\usepackage[OE]{express}

\usepackage[colorlinks=true,urlcolor=blue,linkcolor=black,citecolor=black,breaklinks=true,bookmarks=false]{hyperref}

\begin{document}
\title{Phase matching alters spatial multiphoton processes in dense atomic ensembles}

\author{Adam Leszczy\'nski, Micha\l{} Parniak,\textsuperscript{*} and Wojciech Wasilewski}

\address{Institute of Experimental Physics, Faculty of Physics, University of Warsaw, 02-093 Warsaw, Poland}

\email{\textsuperscript{*}michal.parniak@fuw.edu.pl}
\begin{abstract}
Multiphoton processes in dense atomic vapors such as four-wave mixing
or coherent blue light generation are typically viewed from single-atom perspective. Here we study the surprisingly important effect
of phase matching near two-photon resonances that arises due to spatial
extent of the atomic medium within which the multiphoton process occurs.
The non-unit refractive index of the atomic vapor may inhibit generation
of light in nonlinear processes, significantly shift the efficiency
maxima in frequencies and redirect emitted beam. We present these
effects on an example of four-wave mixing in dense rubidium vapors
in a double-ladder configuration. By deriving a simple theory that
takes into account essential spatial properties of the process, we
give precise predictions and confirm their validity in the experiment.
The model allows us to improve on the geometry of the experiment and
engineer more efficient four-wave mixing.
\end{abstract}

\bibliographystyle{osajnl}

\section{Introduction\label{sec:Section-I}}

Resonant atomic nonlinearities involving higher excited states enable
strong multiphoton interactions, facilitating efficient quantum frequency
conversion \cite{Radnaev2010a,Chaneliere2006}, photon pair generation
\cite{Srivathsan2013a,Willis2011} and single-photon level nonlinearities
\cite{Peyronel2012}. Multiphoton processes also enable selective
and precise control of atomic states \cite{Parniak2016a,Huber2014a}.
In particular, non-degenerate four-wave mixing (4WM) in a four-level
configuration offers extensive
possibilities, such as frequency conversion \cite{Parniak2015,Donvalkar2014,Willis2009a,Becerra2008,Brekke2008}
or coherent blue light (CBL) generation \cite{Meijer2006,Sell2014,Vernier2010,Zibrov2002,Akulshin2009a}. 

In volume atomic vapor, as in bulk nonlinear crystals, phase matching
is required to obtain efficient four-wave mixing. In particular, spatial
arrangement of laser beams must enable the phase-matching condition
\cite{Boyd2008}. Since atomic vapors are dilute in comparison to
solid matter, a unity refractive index for all contributing beams
is often assumed when considering the phase-matching condition, thus making
light propagation effects trivial. This approach corresponds to a
simple single-atom perspective and all phase relations throughout
the ensemble are neglected. Consequently, spatial arrangement of the beams is chosen as if the interaction occurred in vacuum.

These assumptions however, are not correct in the regime where four-wave
mixing is most efficient - close to atomic resonances and in a dense
atomic vapor. Multiple experiments encounter effects that arise due
to phase matching: these effects manifest themselves in quite non-obvious
ways, such as shifting of wave-mixing resonances \cite{Brekke2015}
or strong modifications of measured spectra \cite{Meijer2006}.
Proper identification of important contributions is arduous and although
simulations provide reasonable agreement \cite{Zibrov2002}, little
insight is gained. Only a handful of works consider spatial \cite{Wang2011},
dispersive or wavevector mismatching effects in atomic processes such
as electromagnetically induced transparency \cite{Han2015,Bharti2014238,PhysRevA.85.053837}
or four-wave mixing in a double-$\Lambda$ configuration used to generate
twin-beams \cite{PhysRevA.88.033845}.

Here we show that a vast majority of observed effects may be described
by a simple model of spatial propagation and phase matching. We use
our results to engineer the spatial configuration of the beams to
find that four-wave mixing may be enhanced if phase matching is taken
into account. We also demonstrate direct effects of rapidly-varying
refractive index on wave-mixing resonance position, shape and light emission direction.

\section{Theory\label{sec:TEORYJA}}

\begin{figure}[b]
\begin{centering}
\includegraphics[width=1\textwidth]{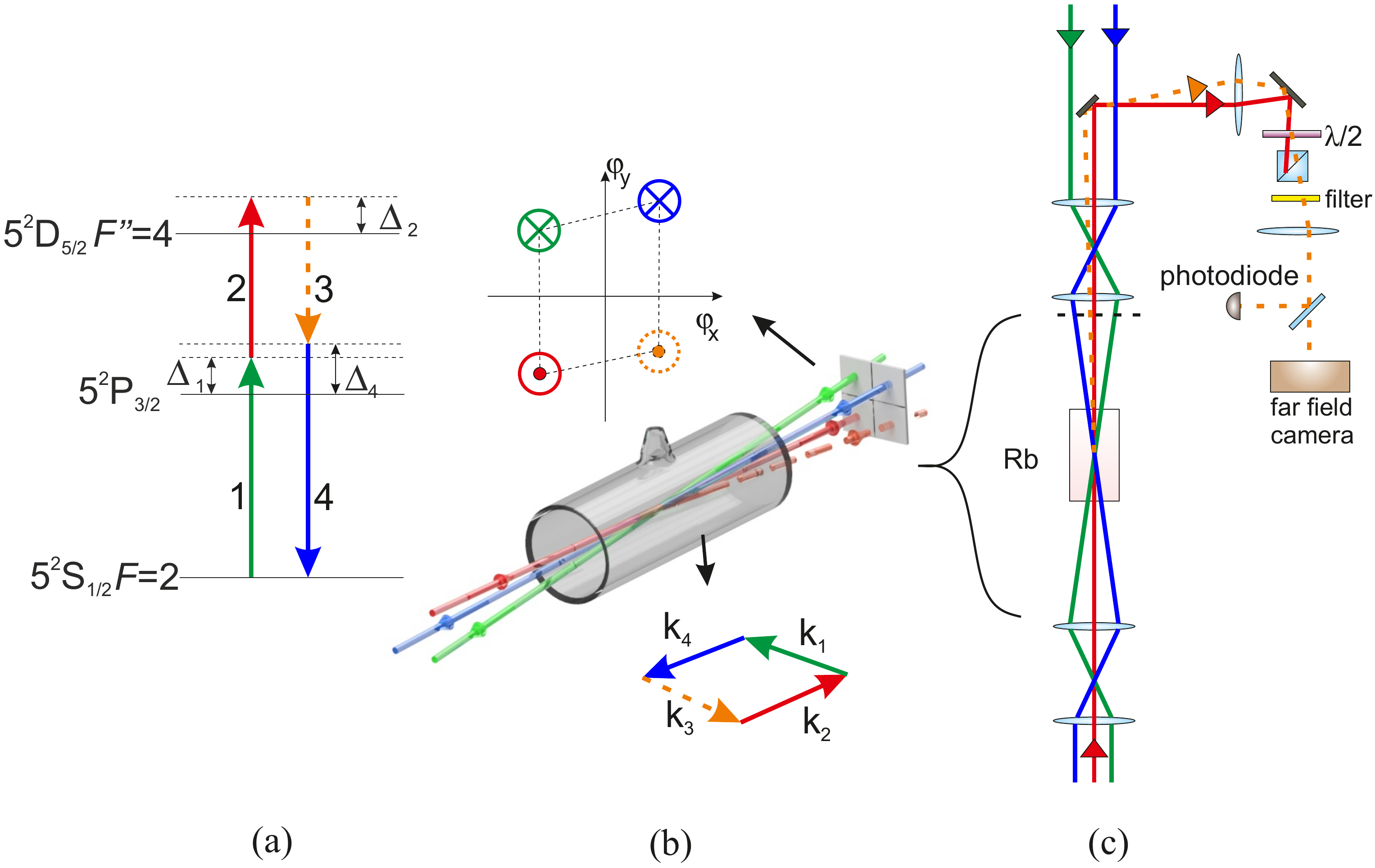}
\par\end{centering}
\caption{ (a) Atomic level double-ladder configuration used in the experiment.
For the ground state $5^{2}\mathrm{S}_{1/2}$ we take one of its hyperfine
components with $F=2$ and for the highest excited state $5^{2}\mathrm{D}_{5/2}$
we take its hyperfine component with $F''=4$. For the intermediate
state $5^{2}\mathrm{P}_{3/2}$ we consider all its hyperfine components,
that lead to multiple wave-mixing paths. (b) Beam directions: three-dimensional
sketch of the laser beams intersecting inside a cell filled with rubidium
vapor. Laser beams marked with green and blue colors have wavelength of
780 nm. Laser beam marked with the red arrow (776 nm) propagates in the opposite
direction. The four-wave mixing signal is marked with the orange dotted line.
The signal is emitted roughly in direction determined by transverse
phase matching, where the sum of transverse components of wavevectors
$\mathbf{\mathbf{k_{1}+\mathbf{k_{2}}}}$ is equal to the respective
$\mathbf{k_{3}}+\mathbf{k_{4}}$ sum. Due to this condition the four
points in the transverse $\mathbf{k}$-plane, corresponding to the
four beams, form a parallelogram. (c) Central part of the experimental
setup. Emitted four-wave mixing signal is filtered using a half-wave
plate ($\lambda/2$), a polarizer and an interference filter and then registered
by with an avalanche photodiode or a CCD camera situated in the far
field with respect to the rubidium vapor cell. \label{fig:schemat}}
\end{figure}

The four-wave mixing signal is generated by the nonlinear polarization
created in the atomic medium by three crossing beams (see Fig.~\ref{fig:schemat}(a)).
We consider a general situation where three incident, non-degenerate
collimated beams (No. 1, 2 and 4) generate a new signal field (beam~3). As presented in Fig. \ref{fig:schemat}(a) the fields are arranged
in the double-ladder configuration being nearly-resonant to atomic
transitions. Beams 1 and 2 take part in two-photon absorption. Beams
3 and 4 take part in two-photon emission, where beam 4 stimulates
the emission \mbox{of the four-wave mixing signal (beam~3).}

To facilitate theoretical description of the process, let us first
consider propagation of a monochromatic signal in a nonlinear atomic
medium with polarization $\mathbf{P}_{NL}(x,y,z)\exp{(i\, k_{0}z-\omega t)}$.
Approximate equation for the electric field envelope $\mathbf{A}$ of
this signal has the following well-known form (see \cite{Boyd2008}):

\begin{equation}
\frac{\partial}{\partial z}\mathbf{A}(x,y,z)=i\frac{\omega^{2}}{2k_{0}c^{2}\epsilon_{0}}\mathbf{P}_{NL}(x,y,z)+\frac{i}{2k_{0}}\Delta_{\perp}\mathbf{A}(x,y,z),
\end{equation}
where $k_{0}\approx\mathrm{\mathrm{{Re}}(}\sqrt{1+\chi(\omega)})\omega/c$,
$\chi(\omega)$ is the linear polarizability, $\omega$ is emitted light frequency and $\Delta_{\perp}=\frac{\partial^{2}}{\partial x^{2}}+\frac{\partial^{2}}{\partial y^{2}}$.
In the transverse Fourier domain ($(x,y)\to(k_{x},k_{y})=\mathbf{k}_{\perp}$)
the equation takes on the following form:

\begin{equation}
\frac{\partial}{\partial z}\widetilde{\mathbf{A}}'(\mathbf{k}_{\perp},z)=i\frac{\omega^{2}}{2k_{0}c^{2}\epsilon_{0}}\widetilde{\mathbf{P}}_{NL}(\mathbf{k}_{\perp},z)\exp{\left(i\, \frac{\mathbf{k}_{\perp}^{2}}{2k_{0}}z\right)},
\end{equation}
where we have used the exponential ansatz for rapidly-varying longitudinal
phase dependance $\widetilde{\mathbf{A}}'=\widetilde{\mathbf{A}}\exp{(i\, \frac{\mathbf{k}_{\perp}^{2}}{2k_{0}}z)}$
and tilde indicates the Fourier transform of a respective field. In the
experiment we start with no four-wave mixing signal at the input $\mathbf{A}(x,y,z=0)=0$,
so the above equation may be integrated to yield the signal field
amplitude at the exit of the atomic medium:

\begin{equation}
\widetilde{\mathbf{A}}(\mathbf{k}_{\perp},L)=i\frac{\omega^{2}}{2k_{0}c^{2}\epsilon_{0}}\int\limits _{0}^{L}\widetilde{\mathbf{P}}_{NL}(\mathbf{k}_{\perp},z)\exp\left(i\, \frac{\mathbf{k}_{\perp}^{2}}{2k_{0}}(z-L)\right)\mathrm{d}z.\label{eq:obw}
\end{equation}
High macroscopic amplitude at the output is generated only when the
spatial oscillations in $z$-direction inside the integral are slow
in comparison with the medium length $L$, and thus the correct phase matching
is required. 

\begin{figure}[b]
\centering{}\includegraphics[width=1\textwidth]{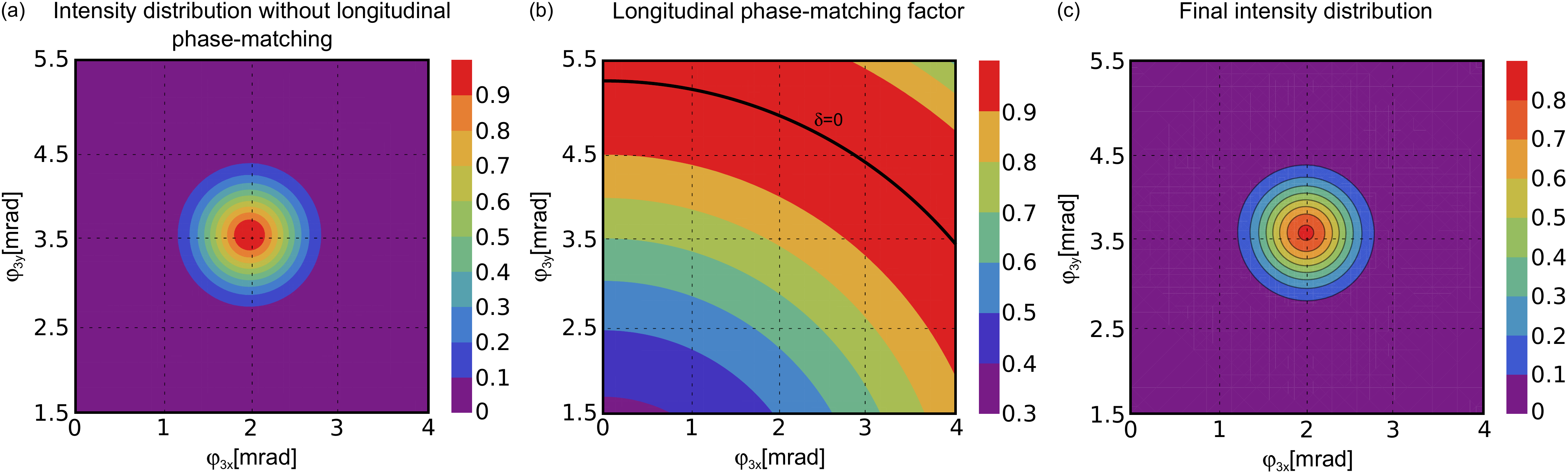}\caption{Theoretical amplitudes of emitted four-wave mixing signal intensity
assuming that 1/$e^{2}$ diameters of incoming beams equal to 400
$\mu$m. Panel (a) portrays the Gaussian intensity distribution without taking into account the longitudinal phase-matching
condition. Panel (b) shows the
value of the phase-matching factor $\mathrm{sinc^{2}}\left(\frac{k_{3}L}{2}\delta\right)$.
The black circle corresponds to a region where the longitudinal phase matching
is perfect. Variations of the refractive index induced by changing
laser detuning cause the radius of this circle to change as well,
which in turn impacts the direction of emission. Panel (c) presents product of (a) and (b), reflecting small change in shape and radial translation. \label{fig:symul}}
\end{figure}

The incoming laser beams indexed by $j=1,2,4$ are assumed to be Gaussian
with equal diameters of $2\sigma$. Their frequencies $\omega_{i}$
determine their wavevector lengths $k_{j}=n_{j}\omega_{j}/c$, with
$n_{j}$ being the refractive index for respective beam. They propagate
at small angles $\boldsymbol{\varphi}_{j}=(\varphi_{jx},\varphi_{jy})$ with
respect to the $z$-axis, so it is convenient to replace wavevectors
with angles $\mathbf{k}_{j\perp}\approx\boldsymbol{\varphi}_{j}k_{j}$,
$k_{iz}\approx(1-|\boldsymbol{\varphi}_{j}|^{2}/2)k_{j}$. The desired final result is the angular distribution of emitted four-wave mixing
signal $\widetilde{A}_{3}(\boldsymbol{\varphi}_{3},L)$ and its total
intensity $I_{3}=\int|\widetilde{A}_{3}(\boldsymbol{\varphi}_{3},L)|^{2}\textrm{d}\boldsymbol{\varphi}_{3}.$
Let us assume that the amplitude of incident beam $j$ is maximal along
the $z$-axis, neglecting shifts of beams centers:

\begin{equation}
A_{j}(x,y,z)=a_{j}\exp{\left(-\frac{x^{2}+y^{2}}{2\sigma^{2}}\right)}\exp{(i\, \mathbf{k}_{j}\cdot\mathbf{r})}.
\end{equation}
 Corresponding nonlinear atomic polarization generated in such a
system is: 

\begin{equation}
P_{NL}(x,y,z)=\chi_{3}^{(3)}A_{1}A_{2}A_{4}^{*}=a_{3}\exp{\left(-\frac{3(x^{2}+y^{2})}{2\sigma^{2}}\right)}\exp{(i\, \mathbf{K}\cdot\mathbf{r})},\label{eq:p4wm}
\end{equation}
where $a_{3}=\chi_{3}^{(3)}a_{1}a_{2}a_{4}^{*}$ and the resulting
polarization wavevector $\mathbf{K}=\mathbf{k}_{1}+\mathbf{k}_{2}-\mathbf{k}_{4}.$
This wavevector is oriented at an angle $\boldsymbol{\Phi}$ to the
$z$-axis, so transverse coordinates are given by 

\begin{equation}
\mathbf{K}{}_{\perp}\approx k_{3}\boldsymbol{\Phi}=k_{1}\boldsymbol{\varphi}_{1}+k_{2}\boldsymbol{\varphi}_{2}-k_{4}\boldsymbol{\varphi}_{4}\label{eq:duzefi}
\end{equation}

Inserting the expression for nonlinear polarization [Eq. (\ref{eq:p4wm})]
with the wavevector given by Eq.~(\ref{eq:duzefi}) to the integral
[Eq.~(\ref{eq:obw})], we obtain the final distribution of the four-wave
mixing signal:
\begin{equation}
\widetilde{A}_{3}(\boldsymbol{\varphi}_{3},L)\sim a_{3}\exp\left(-\frac{k_{3}^{2}\sigma^{2}|\boldsymbol{\varphi}_{3}-\boldsymbol{\Phi}|^{2}}{6}\right)\mathrm{sinc}\left(\frac{k_{3}L}{2}\delta\right),\label{eq:amplituda}
\end{equation}

\begin{equation}
\delta=\underbrace{\frac{k_{1}}{k_{3}}\left(1-\frac{|\boldsymbol{\varphi}_{1}|^{2}}{2}\right)+\frac{k_{2}}{k_{3}}\left(1-\frac{|\boldsymbol{\varphi}_{2}|^{2}}{2}\right)-\frac{k_{4}}{k_{3}}\left(1-\frac{|\boldsymbol{\varphi}_{4}|^{2}}{2}\right)-1}_{-\theta_{3}^{2}/2}+\frac{|\boldsymbol{\varphi}_{3}|^{2}}{2}.\label{eq:delta}
\end{equation}

 The above formula for the distribution consists of two factors: (a)
the Gaussian factor that depends only on transverse variables and
is the result which is obtained if we neglect the longitudinal phase matching
and (b) the $\mathrm{sinc}^{2}\left(k_{3}L\delta/2\right)$ factor
that represents the longitudinal phase-mismatch. Fig. \ref{fig:symul}
shows example theoretical intensity distribution of emitted beam
without taking into account the longitudinal phase-matching condition
(i.e. the Gaussian factor, at the left) and the ring-shaped phase-matching
factor (at the right). Shape of the four-wave mixing intensity distribution
depends on angles and wavevector lengths of incoming beams. The Gaussian
factor is virtually independent of frequencies. Indeed, small changes
of wavevector lengths shift the Gaussian factor by much less than
its width $k_{3}\sigma$, so they can be neglected. However, since
$k_{3}L\approx10^{5}$, we need very small $\delta$ {[}see Eq. (\ref{eq:delta}){]}
for efficient phase matching. Equation (\ref{eq:delta}) can be rewritten
as $\delta=|\boldsymbol\varphi_{3}|^{2}/2-\theta_{3}^{2}/2$ which for $\delta=0$
describes the cone of perfect phase matching. In the transverse plane,
the best phase-matching is achieved on the ring with radius $\theta_{3}$
with angular width of approximately 
\begin{equation}
\label{eq:deltatheta}
\Delta\theta_3=\pi/(2k_{3}L\theta_{3}).
\end{equation}

We now consider changes of the radius $\theta_{3}$ as a function
of incident and emitted beams wavevectors. The derivative over $k_{j}$
is 
\begin{equation}
\frac{\partial\theta_{3}}{\partial k_{j}}\approx\pm\frac{1}{k_{j}\theta_{3}}.
\end{equation}
A typical value we find for $\theta_{3}$ in the experiment is of
the order of $10^{-3}$ rad, so derivative is of the order of $10^{-4}$
m$^{-1}$. Therefore, change of refractive index of the order of $10^{-5}$
causes change of $\theta_{3}$ of the order of $10^{-2}$ rad, much
more than phase-matching ring width. By analogy we can calculate the derivative
over incident beam angles $\varphi_{jx,y}$:
\begin{equation}
\frac{\partial\theta_{3}}{\partial\varphi_{jx,y}}=\pm\frac{k_{j}\varphi_{jx,y}}{k_{3}\theta_{30}}\label{eq:dthdfi}
\end{equation}
Consequently, unless the spatial configuration is properly designed,
the overlap between phase-matching ring and Gaussian factor from Eq.
(\ref{eq:amplituda}) could potentially decrease with any change of
$\boldsymbol{\varphi}_{j}$ and resulting change of the phase-matching
ring radius $\theta_{3}$.

We now consider linear and nonlinear polarizabilities essential to
description of the four-wave mixing process. The third order polarizability
can be obtained through perturbation chain for a specific case we
consider in the experiment \cite{Parniak2015}:
\begin{equation}
\chi_{3}^{(3)}=\frac{Nd_{12}d_{23}d_{23}^{*}d_{14}^{*}}{4\hbar^{3}\widetilde{\Delta}_{1}\widetilde{\Delta}_{2}\widetilde{\Delta}_{4}^{*}},\label{eq:chi3}
\end{equation}
where $\widetilde{\Delta_{j}}=\Delta_{j}+i\Gamma_{j}/2$ , $\Gamma_{j}$
is the decay rate, $d_{ij}$ are dipole moments of transitions between
respective states and $N$ is the atomic concentration. Additionally,
since phase matching plays a crucial role, we take into account dispersion
of refractive index. The linear polarizability for beam 1 is:
\begin{equation}
\chi_{1}=-\frac{N}{\epsilon_{0}}\left(\frac{d_{12}^{2}}{\hbar\widetilde{\Delta}_{1}}-\frac{d_{12}^{2}\Omega_{2}^{2}}{4\hbar\widetilde{\Delta}_{1}^{2}\widetilde{\Delta}_{2}}\right),
\label{eq:chi1}
\end{equation}
where $\Omega_{2}$ is the Rabi frequency for beam 2. The linear polarizability
for beam 4 takes on analogous form. For beam 2 we have the following linear polarizability:
\begin{equation}
\chi_{2}=-\frac{Nd_{23}^{2}\Omega_{1}^{2}}{4\epsilon_{0}\hbar\widetilde{\Delta}_{1}\widetilde{\Delta}_{1}^{*}\widetilde{\Delta}_{2}},\label{eq:chi2}
\end{equation}

and the polarizability for beam 3 will take on an analogous form as well.

To complete our considerations we take the Doppler broadening into
account. All, linear and nonlinear, polarizabilities should be averaged
over the Maxwell velocity distribution \cite{Parniak2015,Mirza2015}:
\begin{equation}
\chi_{j}(\omega_{1,}...,\omega_{4})\to\int\limits _{-\infty}^{\infty}\chi_{j}(\omega_{1}-k_{1}v,...,\omega_{4}-k_{4}v)g(v)\mathrm{d}v,
\end{equation}

\begin{equation}
g(v)=\sqrt{\frac{m}{2\pi k_{b}T}}\exp{\left(-\frac{mv^{2}}{2k_{\mathrm{B}}T}\right)},
\end{equation}
where $m$ is the atomic mass, $k_{\mathrm{B}}$ is Boltzmann's constant and $T$
is temperature. Dominant contribution of the Doppler broadening is
present in single-photon terms, which we take into account as full
Voigt profiles in the above calculation. However, there is also residual
Doppler broadening in the two-photon terms with $\Delta_{2}$. Since
calculation of this term is time-consuming while the broadening is
small, we decided to replace it by additional natural broadening, which is of quite
different shape, but gives similar and consistent results.

Originally our model is created for four-wave mixing in double-ladder configuration, but it is worth to mention that it can be very easily adapted to another schemes like diamond or double-$\Lambda$ configuration and other multi-wave mixing schemes. We only need to change the polarizabilities, which in any case have analogous forms. 
The model is also applicable to the cylindrically-symmetric case, where all beams are co-propagating. Instead of radial shift of emission direction one could observe emission to the cone where the apex angle $\theta_3$ depends on detunings. This regime can be reached if we use beams with sufficiently small diameters $\sigma<1/\theta_3$.

In following sections we describe measurements of four-wave mixing signal characteristics (intensity and direction of emission) as a function of detunings and direction of all incoming beams and we compare it with theoretical prediction of model presented above.

\section{Experimental}

Essential components and ideas of the experiment are presented in
Fig. \ref{fig:schemat}. We use a double-ladder level configuration
in $^{87}$Rb [Fig. \ref{fig:schemat}(a)]. The four-wave mixing signal is
generated on the transition between $5^{2}\mathrm{D}_{5/2}$ and $5^{2}\mathrm{P}_{3/2}$
manifolds. To drive the process we use three lasers: one external-cavity
diode laser (ECDL, number 2) with wavelength of 776 nm and linewidth
approx. 100 kHz and two distributed feedback (DFB) laser diodes with
wavelengths of 780 nm (number 1 and 4) and linewidth of about 1 MHz.
To minimize the two-photon Doppler broadening laser beams 1 and 2
are arranged in the counter-propagating configuration. The 776 nm
laser is stabilized using a commercial wavemeter (HighFinesse WS7).
One of the 780 nm lasers is locked at the vicinity of two-photon absorption
peak using an auxiliary rubidium vapor cell placed in a tunable magnetic
field where we crossed strong circularly polarized 776 nm laser beam
and weak linearly polarized laser 1 beam at 780 nm. By measuring polarization
rotation in circular basis we could generate tunable locking signal,
that controls laser 1 and lock the two lasers at $\Delta_2\neq0$. More details of this method are described in \cite{Parniak2016}. Using the fact that the difference between
frequencies of lasers 1 and 4 is of the order of several GHz, laser
4 is stabilized by beat-note measurement. 

\begin{figure}[b]
\centering{}\includegraphics[width=0.75\textwidth]{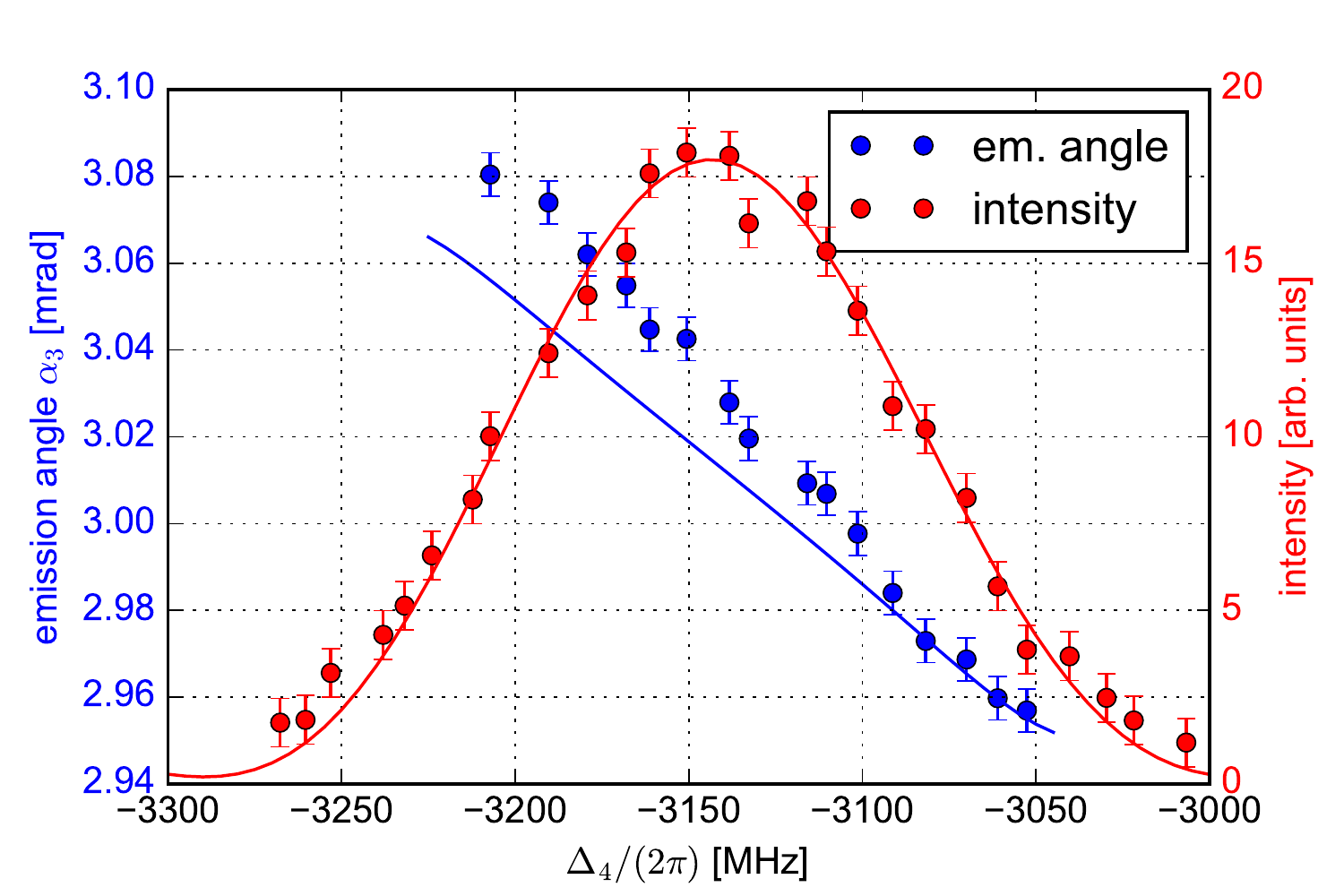}\caption{Dependence of the four-wave mixing signal intensity and emission angle
on detuning $\Delta_{4}$ for $\Delta_{1}/2\pi=-3500$ MHz and $\Delta_{2}=0$
at $T=145\mathrm{^{\circ}C}$. Dots correspond to the experimental
result, while solid lines correspond to theoretical prediction.
\label{fig:przesuwanie}}
\end{figure}

All laser beams intersect in the cell with warm rubidium vapors at
natural abundance and no buffer gas. Fig. \ref{fig:schemat}(b) depicts
the central part of the setup and a telescope used to obtain beams
intersecting at approx. $11$ mrad angle and $1/e^{2}$ diameter of
400 $\mu$m. To eliminate influence of external magnetic fields the cell is placed inside a double $\mu$-metal shielding. Additionally,
the cell is heated using a bifilarly-wound coil to avoid stray magnetic
field from the heater. 

The generated four-wave mixing signal is separated from stray driving
light using a half-wave plate with a polarizing beamsplitter and then
using a band-pass interference filter tilted to transmit light at
776 nm. The signal is registered by an avalanche photodiode (APD)
or a CCD camera situated in the far field of the rubidium cell, allowing
us to measure angular distribution of the emission. A flip-mirror
is used to select either the APD or the CCD camera (Fig. \ref{fig:schemat} (c)).

\section{Results}

\begin{figure}[b]
\centering{}\includegraphics[width=0.75\textwidth]{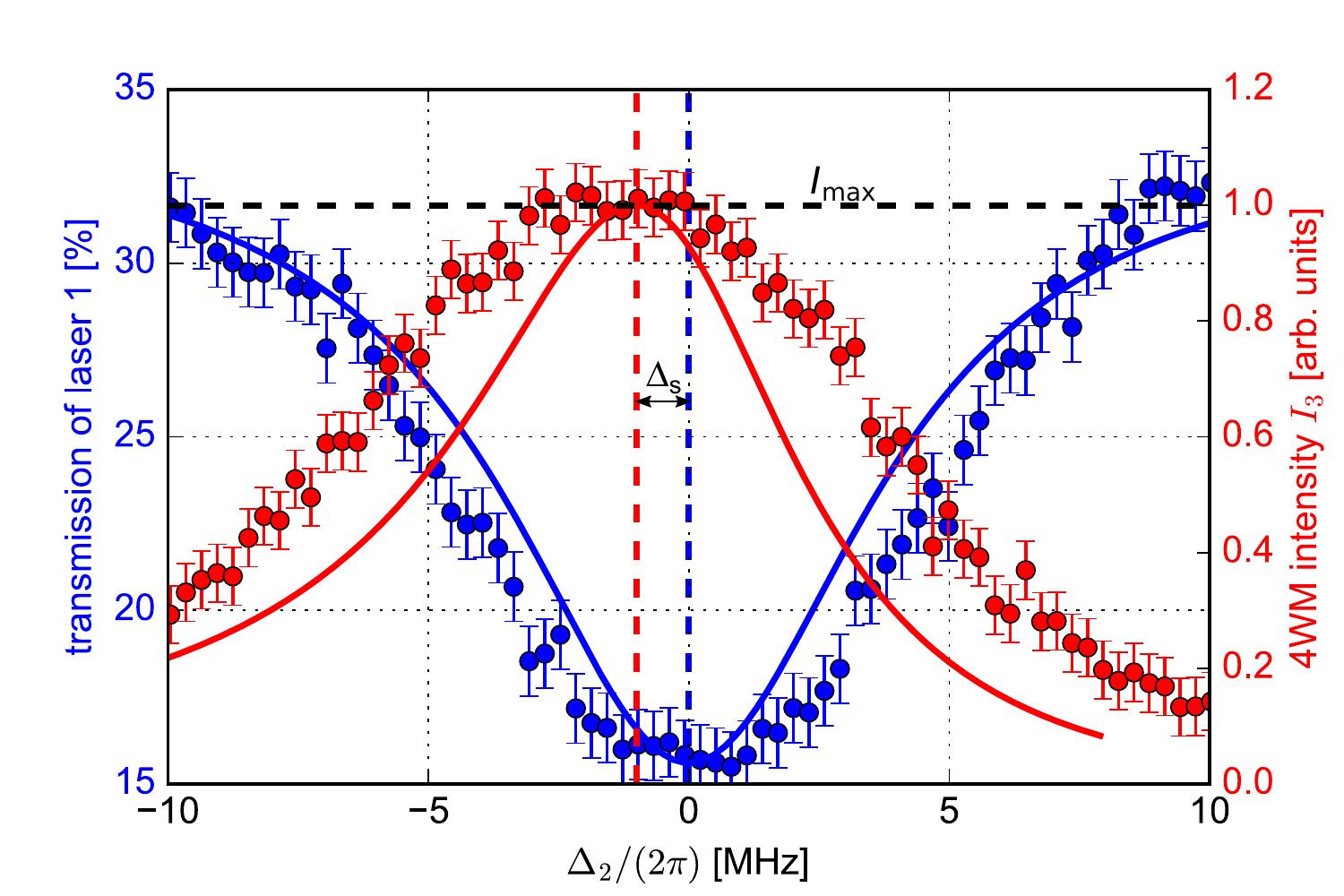}\caption{Transmission profile of laser 1 light (blue curve and points) and
the intensity of four-wave mixing signal (red curve and points) as
a function of two-photon detuning $\Delta_2$ in the vicinity of the
two-photon resonance. Dots correspond to experimental result for $\Delta_{1}/2\pi=-3000$
MHz and $\Delta_{4}/2\pi=-2760$ MHz, while solid lines correspond
to the theoretical prediction. Vertical dotted lines mark maxima of the
four-wave mixing signal intensity and two-photon absorption. The frequency
shift between the two maxima is marked as $\Delta_\mathrm{S}$, while the
maximum intensity of the four-wave mixing signal attained at the resonance
is marked as $I_{\mathrm{{max}}}$ .\label{fig:przes}}
\end{figure}

\begin{figure}[b]
\begin{centering}
\includegraphics[width=0.75\textwidth]{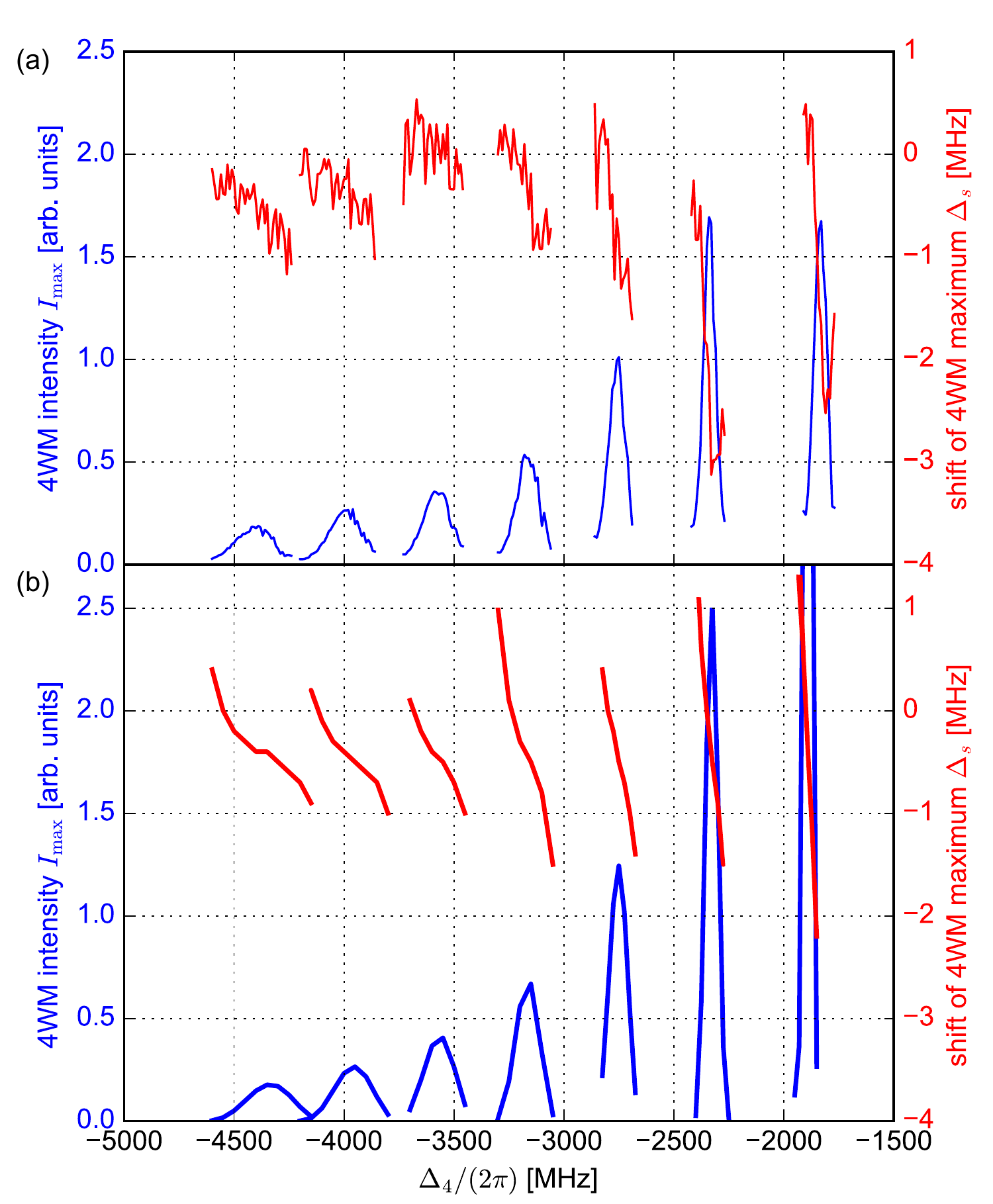} 
\par\end{centering}
\centering{}\caption{Maximal four-wave mixing signal intensities $I_{\mathrm{max}}$ (blue
curves, left axis) and corresponding frequency shifts $\Delta_\mathrm{S}$
(red curves, right axis) for a set of $\Delta_{1}$ and $\Delta_{4}$
detuning. Panels (a) and (b) correspond to experimental and theoretical
results, respectively. Subsequent curves were measured/calculated
for different $\Delta_{1}$ detunings in 500 $2\pi\times$MHz steps
starting from $\Delta_{1}/2\pi=-5000$ MHz on the left.\label{fig:przes_cale}}
\end{figure}

\begin{figure}
\begin{centering}
\includegraphics[width=0.75\textwidth]{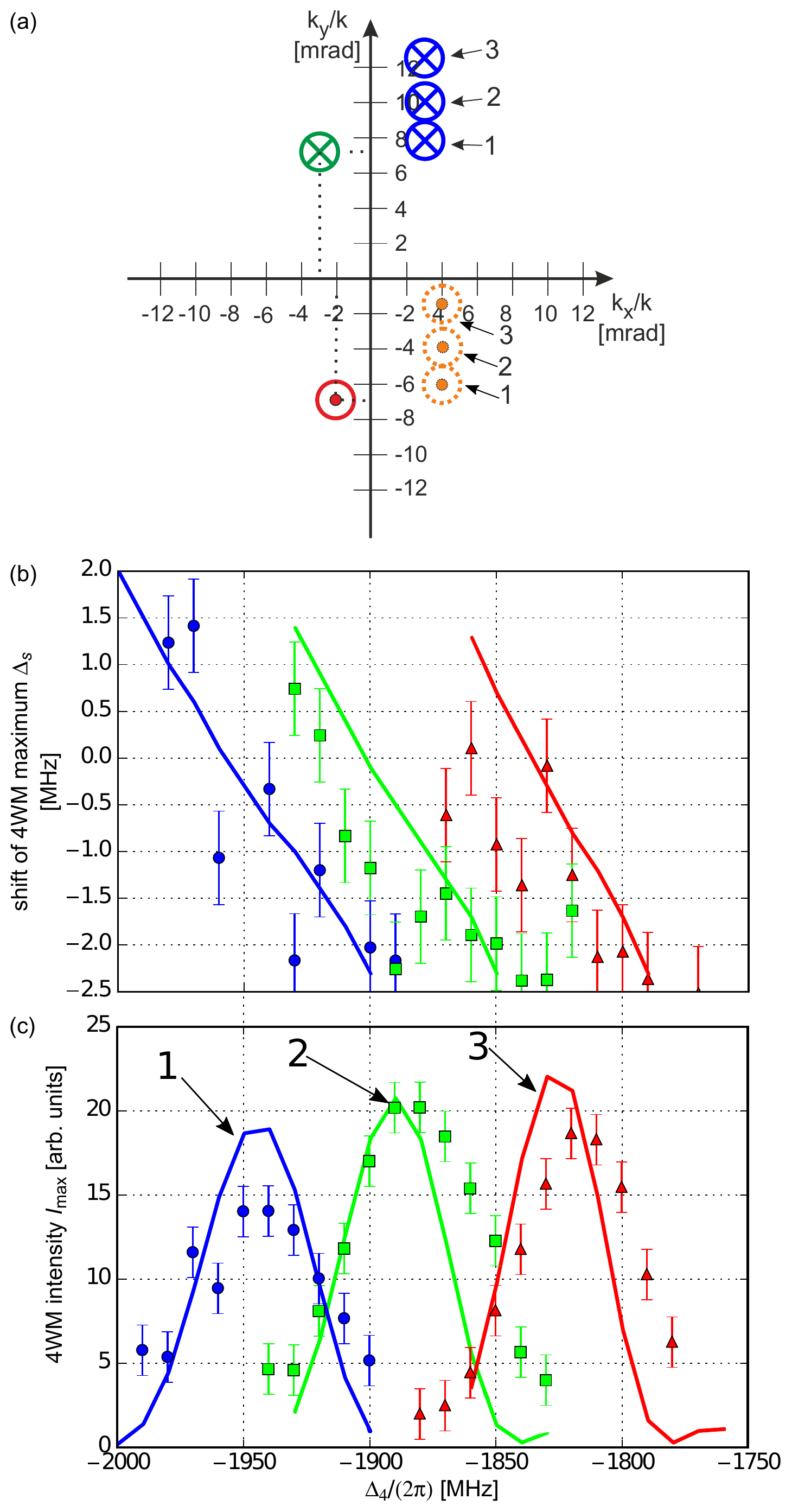}
\par\end{centering}
\centering{}\caption{(a) Tilts of crossing beams to the $z$-axis. Dots and crosses inside
circles correspond to direction to the positive and negative values on the
$z$-axis. (b) Dependence of the four-wave mixing frequency shift $\Delta_\mathrm{S}$ in relation to the two-photon absorption resonance and (c) maximum intensity $I_\mathrm{max}$ on $\Delta_{4}$ detuning.
Subsequent experimental points correspond to measurements at various detuning $\Delta_{4}$ and angles as shown in panel (a) with constant $\Delta_{1}/2\pi=-2000$ MHz.
Dots correspond to experimental result, while solid lines correspond
to theoretical prediction. \label{fig:kat}}
\end{figure}
 In the following section we compare our experimental and theoretical
results to demonstrate the importance of the phase-matching condition
and finally engineer the optimal geometry for the process.

First, we observe the dependence of the average emission angle $\alpha_{3}$
of the four-wave mixing signal on detuning $\Delta_{4}$, while remaining
detunings are kept constant. In particular, driving beam 1 is far
detuned from the resonance ($\Delta_{1}/2\pi=-3500$ MHz), while being
simultaneously kept in the two-photon resonance, i.e. $\Delta_{2}=0$, where the four-wave mixing signal is close to the optimum.
From the theoretical point of view, the phase-matching ring, as described
in Sec. \ref{sec:TEORYJA}, has finite width and overlaps with the
Gaussian factor from Eq.~(\ref{eq:amplituda}) at either side. For that reason, the emission angle $\alpha_{3}$ changes
as a function of detuning (in this case $\Delta_{4}$). In the experiment,
the full signal beam profile at the far field is measured with the
CCD camera and the average emission angle is inferred from the image
after electronic background subtraction. In the theoretical framework this corresponds
to the angle calculated as:

\begin{equation}
\alpha_{3}=\frac{\int\mathrm{d\boldsymbol\varphi_{3}|\boldsymbol\varphi_{3}|}|\tilde{A}_{3}(\boldsymbol\varphi_{3},L)|^{2}}{\int\mathrm{d\boldsymbol\varphi_{3}}|\tilde{A}_{3}(\boldsymbol\varphi_{3},L)|^{2}}.\label{eq:aldadef}
\end{equation}
Measurement results and theoretical predictions results are presented
in Fig. \ref{fig:przesuwanie}. We observe a strong correspondence between
the two, which confirms that the reason of dependence between emission
direction and lasers detunings is indeed dispersion and resulting change in phase matching. 

As long as the change of emission angle is perhaps the most direct
consequence of dispersion and phase matching, another significant
effect is the shift in optimum two-photon detuning $\Delta_{2}$ for
the four-wave mixing signal generation in relation to two-photon absorption
maximum. This effect, sometimes also visible as a strong modification of spectrum, has been observed in several previous works \cite{Zibrov2002,Brekke2015,Sell2014,Parniak2016a},
but its origin has not been studied thoroughly. Simple theory, which neglects
dispersion of refractive indices, predicts that maximum intensity
of measured signal is exactly at the two-photon absorption maximum,
as the third order polarizability $\chi_{3}^{(3)}$ and the two-photon
absorption coefficient proportional to $\chi_{2}$ depends on the two-photon
detuning $\Delta_{2}$ in the very same way {[}see Eqs. (\ref{eq:chi3})
and (\ref{eq:chi2}){]}. In all cases the magnitude of the effect is limited by the two-photon linewidth.

In our model not only large polarizability $\chi_{3}^{(3)}$ is required,
but also overlap between the phase-matching ring and Gaussian factor
from Eq. (\ref{eq:amplituda}). The latter condition is not necessarily
met at $\Delta_{2}=0$. From this reasoning, the frequency shift $\Delta_\mathrm{S}$ of the maximum follows. Similar situation in double-$\Lambda$ configuration was described in \cite{PhysRevA.88.033845} where this effect was also presented as a consequence of dispersion. An exemplary plot of the two-photon absorption and the four-wave mixing
signals as a function of the two-photon detuning are presented in
Fig. \ref{fig:przes}. Note a significant shift $\Delta_\mathrm{S}$ between
the extrema of the two signals, marked with dashed lines. Imperfect
fit to theory is due to residual two-photon Doppler broadening, accounted
for in numerical calculations as a broadened Lorentzian instead of
a full Voigt profile. The shift was measured for multiple values of
detunings $\Delta_{1}$ and $\Delta_{4}$. Experimental results,
together with theoretical prediction, are presented in Fig. \ref{fig:przes_cale}.
For different detunings the system exhibits different dispersion profiles,
and thus we observe different frequency shifts $\Delta_{S}$. They
confirm that dispersion and phase matching accurately explain dependence
between the four-wave mixing signal intensity and laser detunings.

According to Eqs. \eqref{eq:chi3}--\eqref{eq:chi2} all polarizabilities depend linearly on the atomic concentration.  In consequence not only the four-wave mixing becomes more efficient at larger atomic concentrations, but also the dispersive effects become significantly more important. As the atomic concentration strongly depends on cell temperature,
the size of the measured effect will strongly depend on temperature as well, as in e.g \cite{Brekke2015}. 
In our experiments we set the temperature to 145~$^\circ$C, which corresponds to the atomic concentration of the order of $10^{19}$ $\mathrm{m}^{-1}$. This setting allows us to attain large signal intensities together with clearly visible dispersive effects.

Finally, we confirm that as originally intended the geometry of the
experiment may be changed to tailor the four-wave mixing signal properties.
From theory it follows that the overlap between the phase-matching ring
and the Gaussian factor decreases with any change of incident beams
tilts $\boldsymbol\varphi_{1}$, $\boldsymbol\varphi_{2}$ and $\boldsymbol\varphi_{4}$ {[}see Eq. (\ref{eq:dthdfi}){]}.
However, this overlap can be fixed by adjusting detunings. As an
example, we study how the maximum of the four-wave mixing signal
changes as a function of $\Delta_{4}$ detuning if we tilt the incident
beam 3. We performed the measurement for three different tilts presented
in Fig. \ref{fig:kat}(a). Figures \ref{fig:kat}(b) and \ref{fig:kat}(c)
portray obtained experimental and theoretical results for the frequency shift $\Delta_\mathrm{S}$ and maximum intensity $I_\mathrm{max}$, respectively.
We observe that the maximum of the four-wave mixing signal changes quite
significantly. Note, that without taking into account the phase-matching condition we only obtain $1/\Delta_{4}$
dependance for the field amplitude [as in Eq.~\ref{eq:chi3}], which would clearly give a completely
incorrect theoretical prediction in this case. Instead, the proper
choice of geometry is critical when laser 4 is tuned closer or further
from its respective single-photon resonance. These results demonstrate
that our theoretical model may be used to precisely predict the behavior
of the four-wave mixing signal, explaining a variety of intricacies. In
particular, we may predict the optimum geometry for a desired set of
detunings, or vice-versa.

\section{Conclusions}

We have shown that a simple model of four-wave mixing in an atomic medium
that neglects propagation effects could be easily extended by considering
propagation equation with nonlinear polarization to account for phase matching.
In the spatial Fourier domain, the phase-matching condition implies that only
some component wavevectors of the drive beams lead to effective wave-mixing.
Moreover, due to dispersion, the phase-matching conditions strongly
depend on laser detunings. This approach allowed us to explain some
phenomena which were unexplainable earlier, like the frequency shift of
the four-wave mixing signal maximum in relation to two-photon absorption
maximum \cite{Zibrov2002,Brekke2015,Sell2014,Parniak2016a}. The
most direct consequence of dispersion seems to be the change of the
four-wave mixing signal emission angle as a function of laser detunings. 

These results show that our theoretical description is much more robust
than an atomic model neglecting the influence of dispersion on phase matching
in atomic medium and allows precise predictions of the four-wave mixing
signal behavior. Moreover, we have shown that our model facilitates
proper choice of geometry and consequently stronger signal may be
obtained. Apart from providing a recipe for engineering of effective interaction,
we believe that our results may also help study other subtle effects in
four-wave mixing \cite{DeMelo2015}, where multiple shift effects might contribute to the total frequency shift. 

Finally, we note that our approach
is generally valid for a variety of systems - one needs to simply
supply appropriate expressions for linear and nonlinear susceptibilities.
In particular, effects treated in this work become even more critical in
high-density ensembles, that facilitate strong light-matter interactions
due to collective effects \cite{1367-2630-15-8-085027,Srivathsan2013a},
in waveguides that support modes with various longitudinal wavevectors
\cite{Donvalkar2014}, in cavity-enhanced processes \cite{Offer:16}
or finally in schemes that rely on phase-matching control, such as
multimode quantum memories \cite{Mazelanik:16}. Our model is also applicable to the case of spontaneous four-wave mixing \cite{Srivathsan2013a}: if we post-select on one of the fiber-coupled photons, we may predict the correlated emission direction. With further extensions our theory may also be used to model the full biphoton joint spectral/spatial properties.

If the four-wave
mixing signal is weak as in quantum light-atom interfaces \cite{Parniak2016a,Radnaev2010a}
or processes involving Rydberg states \cite{Brekke2008,DeMelo2014a,Magno2001}
proper choice of geometry may be even more critical and yield a difference
between having a strong signal or not obtaining the four-wave mixing
signal at all. 

\section*{Funding}
The project was financed by the Polish Ministry of Science and Higher
Education ``Diamentowy Grant'' Project No. DI2013 011943 and by
the National Science Centre (Poland) Grants No. 2015/16/S/ST2/00424 and 2015/17/D/ST2/03471. 

\section*{Acknowledgments}
We acknowledge generous
support of T.~Stacewicz, K.~Banaszek and R.~\L{}apkiewicz as well as careful proofreading
of the manuscript by M. D\k{a}browski.

\end{document}